\documentstyle[epsf]{l-aa}
\begin{document}
\thesaurus{
              03         
              (03.13.6;  %
               11.17.4 0957+561A,B;
               12.07.1)  %
             }
\title{ Time delay controversy on QSO 0957+561 not yet decided } 
\author{ J.~Pelt\inst{1}\and W.~Hoff\inst{2}\and R.~Kayser\inst{2}\and
 S.~Refsdal\inst{2}\and T.~Schramm\inst{2} }
\offprints{ R.~Kayser }
\institute{Tartu Astrophysical Observatory,
              EE2444 T\~{o}ravere Estonia \and 
           Hamburger Sternwarte,
              Gojenbergsweg 112,
              D-21029 Hamburg-Bergedorf, Germany
             }
\date{received <date>; accepted <date> }
\maketitle
\markboth{ J.~Pelt et al.: Time delay controversy on QSO 0957+561
not yet decided}{}

\begin{abstract}
We present a new analysis of 
previously published 
optical and radio data sets of the
gravitationally lensed quasar 0957+561 A,B with the aim of determining
the time delay between its two images. We use a non-parametric estimate
of the dispersion of the combined data set where, however, we only make use
of alternating neighbours in order to avoid windowing effects. From the
optical data a time delay of (415 $\pm$ 32) days and from 
the radio data a delay of (409 $\pm$ 23) days is suggested.
We demonstrate a considerable sensitivity of different delay estimation
procedures against the removal of only a few 
observational data points or against
smoothing or detrending of the original data sets.
The radio data give us
formally
a slightly more precise value for the time delay than the optical data. 
Also, due to the
lack of windowing effects, the result obtained for the radio data can
be considered as
somewhat
more reliable than the delay determined from the optical data. 

\end{abstract}
\section{Introduction}
The time delay between the images of a gravitationally lensed object is
of great astrophysical interest since it may be used to determine the 
Hubble parameter as well as the mass of the lens (Refsdal \cite{Refsdala},
\cite{Refsdalb}, 
Borgeest \cite{Borgeest}, Falco et al.~\cite{Falco91b}).
The double quasar 0957+561 A,B 
(Walsh et al.~\cite{Walsh}, Young
\cite{Young}, Falco \cite{Falco92}) is up to now the 
only gravitational lens system
for which serious attempts have been made to determine the time delay
$\tau$ between its images. However, the results are still controversial,
with suggested time delays of between 376 and 657 days (Florentin-Nielsen
\cite{Florentin}, Schild \& Cholfin \cite{Schild86}, 
Gondhalekar \cite{Gondh}, Gorenstein et al. \cite{Goren},
Leh\'ar et al. \cite{Lehar89}, Vanderriest \cite{Van89}, 
Schild \cite{Schild90}, 
Falco et al. \cite{Falco91},
Leh\'ar et al. \cite{Lehar92}, Roberts et al. \cite{Roberts}, 
Press et al. \cite{Pressa},
\cite{Pressb}, Beskin
\& Oknyanskij \cite{Beskin}).
Using an elaborate statistical method (see Rybicki et al. \cite{Rybicki})
Press et al. (\cite{Pressa}, below PRHa),
recently obtained a value of ($536\pm 12$) days from the optical
lightcurve, which is seemingly
in good agreement with the value obtained from the radio data
by the same authors (Press et al. \cite{Pressb}, below PRHb).
They conclude that ``delays less than about 475 days are strongly
excluded''.
It is also often 
claimed (Leh\'ar \cite{Lehar92}, Roberts et al. \cite{Roberts}, 
PRHb)
that the radio data decisively exclude a 
delay of around 415 days, a value favoured
by other authors (Vanderriest et al. \cite{Van89}, Schild 1990).

The time delay obtained by PRHa is quite close to 1.5 years, a
value for which windowing effects due to the uneven sampling of the 
optical data
are expected to be strongest (Vanderriest et al. \cite{Van92}).

We here present results of a careful re-analysis of the same
data sets as used in PRHa and PRHb
using simple explorative type statistical methods. The main emphasis
of our work is the evaluation of two competing 
hypothetical time delays: 415 and 536 days.

For the combined data set generated
from the data of image A and the 
data of image B, time shifted by $\tau$, 
we basically estimate the dispersion $D^2$ of the scatter
around the unknown mean curve.
The true time delay
between the images should then show up as a minimum in the dispersion
spectrum $D^2(\tau)$. It is our aim to determine the dispersion of the 
combined data due to the alternation between the two light curves, and not
the dispersion within each lightcurve. In order to avoid strong windowing
effects we therefore take into account only alternating neighbouring
pairs in the combined data set, i.e. only pairs where one point is
from A and the other one from B, respectively. It should be noted that for
$\tau=(n+0.5)$ years (where $n$ is integer) the dispersion $D^2(\tau)$ would
otherwise be strongly dominated by the inner dispersion of the two original
light curves, naturally leading to pronounced minima.

In Section 2 of the paper we introduce the 
basic statistical techniques used.
In Section 3 we present the 
results of a detailed analysis of the light curves
of the double quasar 0957+561 A,B. 
We here use the same data sets as
PRHa and PRHb to allow a comparison of the results and to
understand the problems leading to the different results. 
In the concluding part
we try to summarize the work done on time delay estimation up to now 
and
to delineate some inner difficulties of the problem under study.
Technical details of the nonparametric dispersion estimation are
treated in the Appendix.
\section{Data analysis methods}  
\subsection{The optimal prediction}
In the papers PRHa, PRHb and 
Rybicki et al. (\cite{Rybicki}) a rather complex procedure
for an optimal interpolation of  the lightcurves
was developed. 
The procedure itself is
practically the same as  used in geophysical research under the
name of {\em kriging} (Cabrera et al. \cite{Cabrera}).
The use of sound statistical arguments 
and validation of the results by Monte-Carlo simulations allowed
the authors to draw conclusions on the probable time delay 
with a (formally) 
high level of confidence. Since our analysis did not fully confirm
these results, we reimplemented the 
procedures used in PRHa and PRHb and
used some simple procedures to check the
inner consistency of the results. 
\begin{figure}
\leavevmode\epsfbox{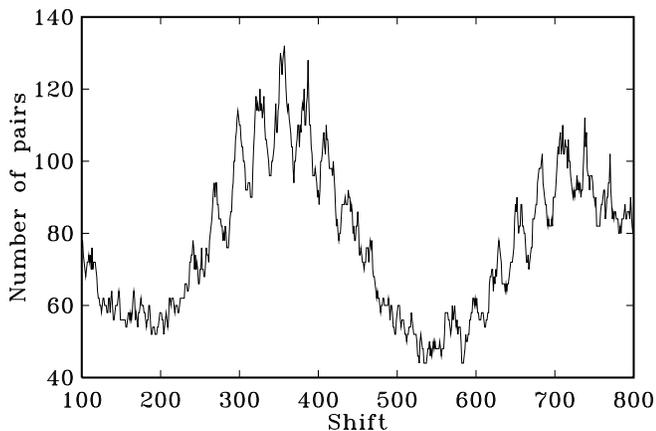}
\caption[ ]{Window function for optical data}
\label{fig1}
\end{figure}
We started from a very simple computation of the  number of mixed data
pairs in the combined sequence (data set B shifted by $\tau$ 
coalesced with data set A). In Fig.~\ref{fig1} the number
of  neighbouring data pairs of mixed type 
(one from data set A, one from 
data set B) is plotted against the trial time delays. 
This simple statistic measures well the total overlap between the two
data sets for different shifts in time of the B component.
Inside of the  range of probable 
delays quoted in PRHa
two strong minima at $\tau=527$ days,$N_{A,B}=43$ and at $\tau=535$ days,
$N_{A,B}=43$
are found.
There is  a  strong similarity between the window 
function plot in  PRHa and our simple statistic. It is
quite obvious (and was noticed also by PRHa) 
that the coincidence of minima in the window function
and in the $\chi^2$ spectrum in PRHa is a rather serious signal
about the complexity of problems one encounters with the
lightcurves of QSO 0957+561. 

After reimplementation of the PRHa procedures we 
carried out the following simple experiment. We divided the data points
of the combined data set (computed with $\tau=536$ days) into two subsets:
relevant and irrelevant. Into the first subset we included observations
which had at least one nearest neighbour from alternate original 
data sets, and into the second subset we included observations for which 
both
neighbours had the same origin as the data point under consideration.
After computation of the optimally approximated (predicted) 
values for all data points,
we were now able to compute separately prediction errors for relevant
and irrelevant observational points. {\em We found that the dispersion
for relevant points was by a factor of five larger than that for irrelevant
points}. 

This result lead us to another interesting 
test. We repeated Monte-Carlo
simulation runs with different time delay values 
exactly as described in PRHa
and used our data set subdivision procedure for each of the 
generated artificial data
sets. 
\begin{figure}[t]
\leavevmode\epsfbox{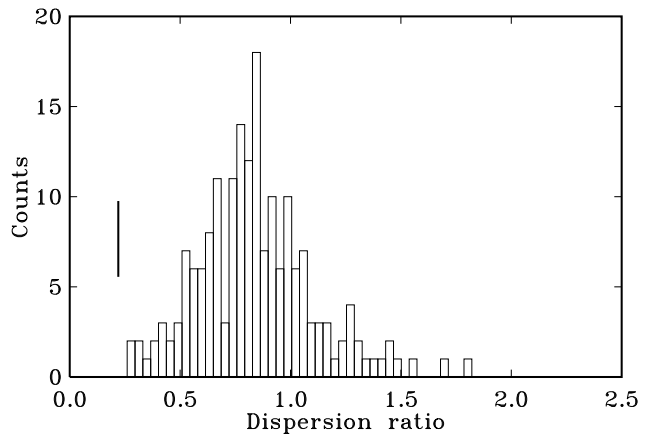}
\caption[ ]{Distribution of dispersion ratios for artificial 
light curves generated by Monte-Carlo
method. The value obtained from the original optical 
data is marked with a dash.}
\label{fig2}
\end{figure} 
In Fig.~\ref{fig2} the resulting distribution of
dispersion ratios is shown. The mean ratio is 
slightly less than unity. This fact can
be explained in the following way: the overlap regions for the two original
data sets are populated in the mean two times more densely than
regions where only one data set is seen. For the sparsely populated
regions there is 
a tendency for the optimal prediction procedure to put the prediction
curve through the observed points
and consequently in these regions the dispersion for prediction errors
is as a rule less than for overlapped regions.
However, as can be seen
from the distribution diagram, the typical decrease is by far not as severe as
for the
original data set (marked in Fig.~\ref{fig2}). 

The conclusion is
discouraging: the  Monte-Carlo procedure
proposed in PRHa, which
was considered as {\it experimentum crucis} for their time delay  
computations does not model the actual data set at hand, but {\em it 
proves only that for data sets with certain regular statistical properties
their procedure works quite well}.  The real data set is unfortunately
more complex, this is probably caused by microlensing.

Our negative experience with the optimal prediction procedure
lead us to the conclusion 
that to understand all the intricacies involved with the actual observations 
we must seek some extremely simple statistics whose use is not
shadowed by the complexities of the mathematical procedures involved. Our
aim was not so much to prove that one or another delay value is right
or wrong, but to understand how 
the different values are ``cooked'' from the data set.

\subsection{Use of nonparametric dispersion estimates for shift analysis}
Let $A_i=g(t_i)+\epsilon_{\rm A}(t_i), i=1,2,\dots,N_A$ and 
$B_j=\big(g(t_j+\tau)-a\big)+\epsilon_{\rm B}(t_j),j=1,2,\dots,N_B$ 
be two time series
which are different measurements 
of the original light curve $g(t)$. It is assumed
that for channel B the observations $B_j$ are amplified with an
unknown amplification ratio $a$.
Observed values contain
also observational errors $\epsilon_{\rm A}(t_i),
\epsilon_{\rm B}(t_i)$ for which we
may have certain dispersion estimates $\delta_{\rm A}^2(t_i),
\delta_{\rm B}^2(t_j)$ or
corresponding statistical weights 
$W_{\rm A}(t_i),W_{\rm B}(t_j)$. Our task is to
estimate the time shift parameter $\tau$ from observed time series.

For every fixed set of values $\tau,a$ 
we can construct a combined data set
from both observed series
\begin{equation}
C_k(t_k)=\cases{{A_i},& if $t_k=t_i$,\cr
                {B_j+a},& if $t_k=t_j+\tau$ \cr}
\end{equation}
and estimate its dispersion $D^2(\tau,a)$ 
with the simple nonparametric method described in the Appendix. 
The {\it dispersion spectrum} is now
defined as:
\begin{equation}
D^2(\tau)=\min_{a} D^2(\tau,a).
\end{equation}

Statistical weights can be introduced into our scheme by using
the usual combined weights for every pair of observations.
If the statistical weight of  channel A is $W_i$ and that
of  channel B is 
$W_j$ then the dispersion for the difference $A_i-(B_j+a)$ is equal to
\begin{equation} 
{1\over W_i}+{1\over W_j},
\end{equation}
and the squared difference of the two observations
must be consequently included with weights
\begin{equation}
\label{weight}
W_{i,j}={W_i W_j\over W_i+W_j}.
\end{equation}
A correct normalization is obtained by 
dividing the corresponding sums of squared differences by the total
sum of weights.

By choosing different selection windows (see Appendix) 
we can now compute different 
statistics and compute corresponding dispersion spectra. In 
several numerical experiments we tried to understand 
the behaviour of many different variants depending on features in the data
sets at hand. However, to present our results we use only one, and
by the way the simplest algorithm: 
only neighbouring data points are considered
as near enough, to be included in the cross-sums. 
We even do not specify an upper limit $\delta$ for a maximum allowed time 
separation between two sequential points.
This approach allows
us to avoid {\it any free parameters} in our procedures, and makes the
calculations simple enough to be well reproduced by other researchers.

Correspondingly,   
we use in the following analysis only two simple statistics.
The first one
\begin{equation}
\label{firststat}
D_{\rm all}^2=\min_a {\sum_{k=1}^{K-1}W_{k,k+1} (C_{k+1}-C_k)^2\over  
2\sum_{k=1}^{K-1}W_{k,k+1}}
\end{equation}
measures 
the general dispersion of the
combined data set. The second statistic
\begin{equation}
\label{secondstat}
D_{\rm A,B}^2=\min_a {\sum_{k=1}^{K-1}W_{k,k+1} G_k (C_{k+1}-C_k)^2\over  
2\sum_{k=1}^{K-1}W_{k,k+1} G_k}
\end{equation}
where $G_k=1$ only when $C_{k+1}$ and $C_k$ are from different
data sets and $G_k=0$ otherwise, measures the dispersion in the
overlap areas of the combined light curve.
\subsection{Computation of error bars for time delays}
To get an idea about error bars for the minima in the dispersion spectra we
used a simple bootstrap procedure. 
We applied a 7-point median filter to
smooth the combined light curve and reshuffled the corresponding residuals
1000 times to generate bootstrap samples. The full procedure turned out
to be not
very dependent on the median filter length, we therefore present
only results for the particular value of 7. 

It must be stressed that
the somewhat arbitrary value for the filter length was used only to estimate
mean square errors of the computed values of the minima, the minima themselves
do not depend on any prechoosen numerical constant. 

It is possible to get
much more narrow limits to the computed shift estimates by using
longer smoothing filters, but we wanted to be as conservative as possible.

\subsection{Influential data segments}
To demonstrate the effect of removing short segments of data from the analysis
we used the following simple procedure. For any starting index value $l$
in the row of observations, for any skip length $m$ and for 
any pair of hypothetical shift values
$S_1,S_2$ which are interesting for us we can compute
the statistic ({\it gain} of hypothesis $\tau=S_1$ against $\tau=S_2$)
\begin{equation}
\label{gain}
I(l,m,S_1,S_2)=D_{\rm A,B}^2(l,m,S_1)-D_{\rm A,B}^2(l,m,S_2)
\end{equation}
where the dispersions are computed for data sets with skips (start
of data skip from $l$, length of skipped subset $m$ points). 
In various experiments we used different skipping schemes: sometimes
we skipped observing points for both channels, sometimes only from
one channel.
The statistic $I(l,m,S_1,S_2)$
allowed us to compare the influence of skipped values for two different
hypothetical time delays.
Maxima and minima in $I(l,m,S_1,S_2)$ 
indicate skips which favor hypothesis
$S_2$ and hypothesis $S_1$, respectively. 

This kind of approach is quite
common in standard regression analysis where robustness of the predicted curve
is tested against removal of a small number of observed values (outliers).
In our case we assume that the correct estimate for the time delay 
should not
depend strongly on a small number of observations. If it
does, we conclude that some of the data points are real outliers, or
that the available data sets do not contain enough information to
discriminate between the two hypotheses.

\section{Analysis of observed data sets}
\subsection{Optical data}
\begin{figure}[t]
\leavevmode\epsfbox{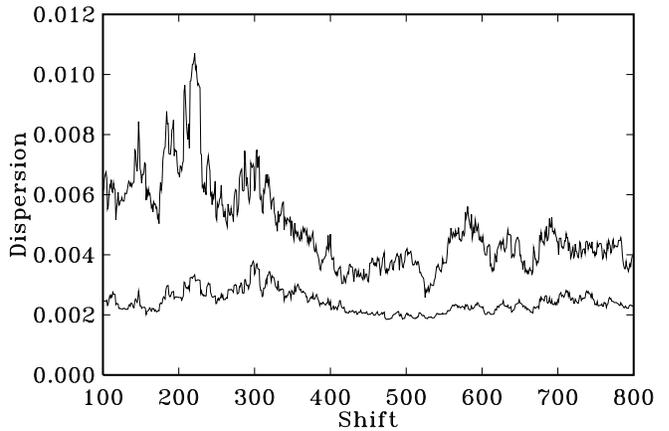}
\caption[ ]{Dispersion spectra for original optical data. Upper curve is
$D^2_{\rm A,B}(\tau)$, lower curve is $D^2_{\rm all}(\tau)$}
\label{fig3}
\end{figure}

We used the same data as PRHa, i.e.~the optical data published by
Vanderriest et al.~(\cite{Van89}).
The $D_{\rm all}^2$ (Eq.~\ref{firststat}) and $D_{\rm A,B}^2$ 
(Eq.~\ref{secondstat}) statistics for the 
original optical data
set are depicted in Fig.~\ref{fig3}. The 
strongest local minima in the $D_{\rm A,B}^2$
spectrum is at a shift of $525$ days and bootstrap calculations give us
a standard error value of $42$ days, so that our result is in good agreement
with the value given in PRHa. 
It is interesting to note that the mean shift for the bootstrap
trials occurred to be biased to a shift of $534$ days. 
There is an additional depression around
$415$ days which is, however, somewhat weaker than the main ``peak''.
It is important to note that the $D_{\rm A,B}^2$ spectrum stays 
{\em everywhere}
well above the $D_{\rm all}^2$ spectrum, i.e.~the dispersion
in the overlap region is systematically higher than in the intervening
regions. We therefore conclude that there must be a 
certain amount of additional mutual variability between two light curves
(maybe due to microlensing), so that the  model of shifted
but otherwise similar images, which has been used here and in PRHa,
is oversimplified.

\begin{figure}[t]
\leavevmode\epsfbox{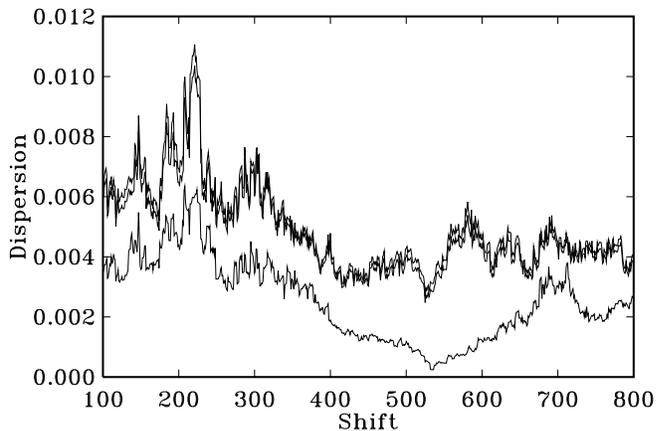}
\caption[ ]{Dispersion spectra for original and artificial data
sets.The two upper curves are $D^2_{\rm A,B}(\tau)\pm 
{\bf D}(\tau)^{1\over 2}$ for the original data, the lower curve is
$D^2_{\rm A,B}(\tau)$ for the artificial data}
\label{fig4}
\end{figure}

To get an idea about the form of the real depression in the dispersion
spectra we proceed with a simple numerical experiment. 
We combined the A and B datasets with the best shift and the best $a$ value,
filtered the combined data set with a 7-point median filter, and
decompose the filtered set back into two subsets. 
The spectra for the real data set and for the 
artificial data set (with a delay of 536 days) 
 are given in Fig.~\ref{fig4}. 
The spectrum for the real data set is
depicted in the two nearby curves. The upper 
curve is the dispersion
spectrum {\it plus} the estimated subsampling error
for the particular
shift value and the lower curve is the dispersion spectrum 
{\it minus} the subsampling
error estimate. Thus, the difference between the two curves demonstrates
the uncertainty due to the subsampling. The minimum value in
the spectrum for the artificial set (which is $D_{\rm trend}^2$ if we use
the notation from the Appendix, see Eq.~(\ref{trend})) 
is comparable to the largest difference
between the upper and the lower limit for
the original spectra. It is therefore clear
that our dispersion estimation scheme works here 
(at least in the region of the
main minimum) nearly optimally.
It can be seen that the 
``line profiles'' of the real and the artificial spectra differ significantly.
The feature in the real spectrum is more fluctuation like than the wide and
systematic depression in the artificial spectrum. 

We then investigated the robustness of our first solution against
skipping short segments from both light curves. The strong motivation
to do this kind of analysis stems from the remark of PRHa that the ``feature
at epoch 6200'' can bias estimation of the shift value towards $415$
days. 

\begin{figure}[t]
\leavevmode\epsfbox{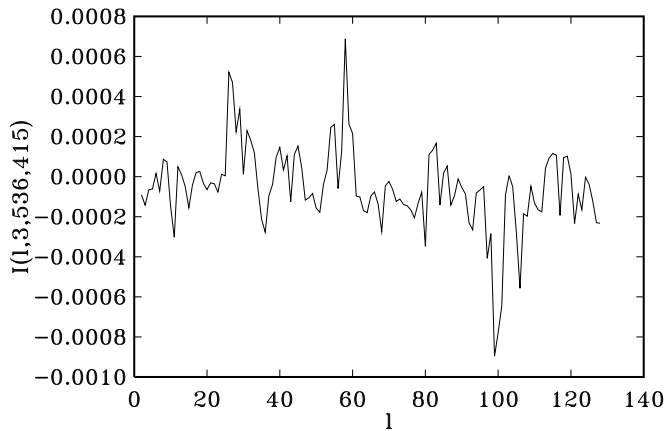}
\caption[ ]{Gain of hypothesis $\tau=536$ days against hypothesis
$\tau=415$ days for different locations of the three point gap in the original
data set}
\label{fig5}
\end{figure}

\begin{figure}[t]
\leavevmode\epsfbox{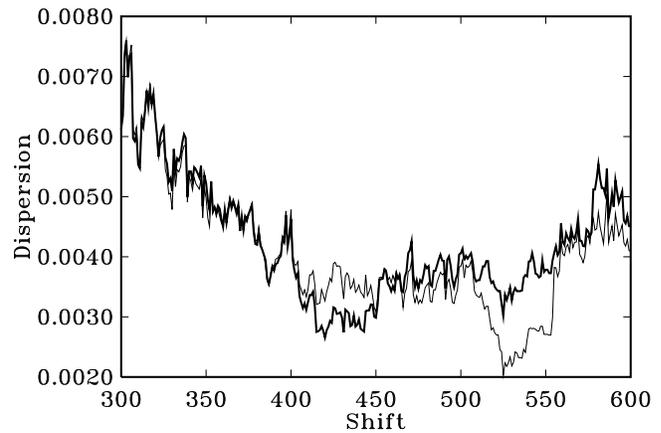}
\caption[ ]{Comparison of the two dispersion spectra obtained by
different skips from the data sets. The three point gaps started from
$l=58$ for the bold curve and from $l=99$ for the normal curve}
\label{fig6}
\end{figure}

In Fig.~\ref{fig5} the results of a typical sensitivity
analysis are shown. The gain $I(l,3,536,415)$ (Eq.~\ref{gain}) 
of hypothesis $\tau=536$ 
days against $\tau=415$ days is plotted
against the starting index of a three point skip in the original
data set. 
In Fig.~\ref{fig6} the corresponding spectra are plotted: one
with the most favorable (for  hypothesis $\tau=536$ days,
minimum at $l=99$) skip, 
and one with the 
most unfavourable (maximum at $l=58$). It is seen that
the feature in the region of $536$ days undergoes a rather dramatic change, 
whereas
the feature around $415$ days remains relatively stable. By skipping only
{\it three consecutive time points} 
(the A and B channels are observed simultaneously) we
can strongly emphasize the $536$ feature, or remove it completely! There
is no need to say that for longer gaps the effect of skips 
is even more drastic. 

From these experiments we concluded that it is reasonable to investigate
the high frequency behaviour of the optical data sets. 
Since our investigation
is of explorative type, we  postulated freely (at least for the moment) that
both light curves are corrupted by 
independent low frequency components (which
can be physically interpreted as microlensing). We assumed that these
components can
be modeled by low degree polynomials. 
\begin{figure}[t]
\leavevmode\epsfbox{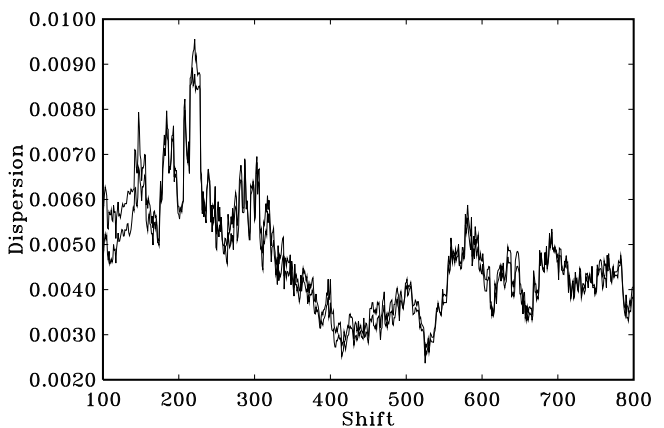}
\caption[ ]{Dispersion spectra for the detrended data sets.
On the plot the results for degrees $2$ and $4$ are depicted.}
\label{fig7}
\end{figure}
\begin{figure}[t]
\leavevmode\epsfbox{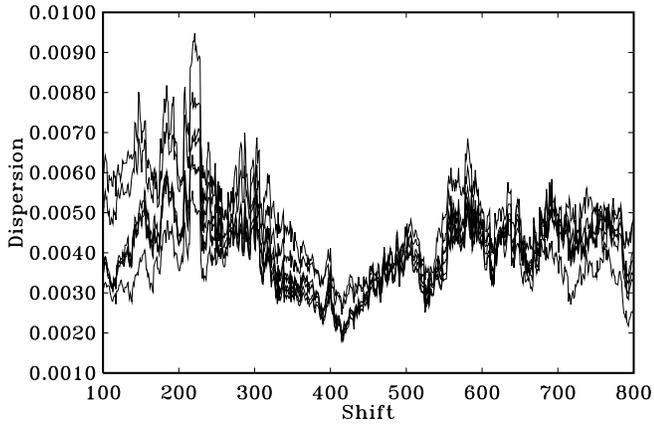}
\caption[ ]{Dispersion spectra for the detrended data sets.
On the plot the results for degrees $1,3,5,6,7,8,9$ are depicted.}
\label{fig22}
\end{figure}
In Figs.~\ref{fig7}  and \ref{fig22} the 
$D_{\rm A,B}^2$ spectra for data sets which were obtained by removing 
polynomial trends with degrees $1-9$ from the original data sets are shown.  
The feature around $415$ days is much stronger than the 
feature around $536$ days for degrees 1,3,5,6,7,8,9. For degrees 2 and
4 (see Fig.~\ref{fig7}) the minima for both 
hypothetical shifts are comparable. 
It is evident that starting from degree 5 the behaviour of dispersion spectra
is quite similar. For our further
presentations we arbitrarily choose polynomials 
with degree $9$ and in the following
discussion we use as detrended data sets the corresponding least squares
fit residuals. The computations with other degrees (from 5 to 8) 
gave essentially the same
results.
\begin{figure}[t]
\leavevmode\epsfbox{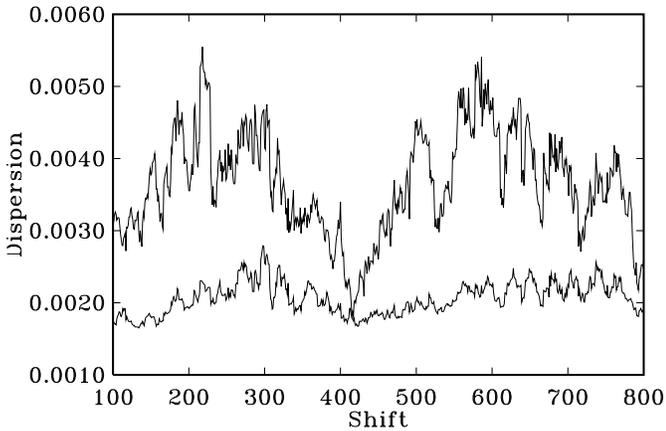}
\caption[ ]{Dispersion spectra for the data sets
detrended by polynomials of degree 9. The upper curve
is $D^2_{\rm A,B}(\tau)$ and the lower curve is $D^2_{\rm all}(\tau)$.}
\label{fig8}
\end{figure}
In Fig.~\ref{fig8} the $D_{\rm all}^2$ and $D_{\rm A,B}^2$ spectra for 
the detrended data
are depicted. 

It is worth to note that  the minimum of
$D_{\rm A,B}^2$ now nearly coincides with the 
value of $D_{\rm all}^2$ for the same
shift. Correspondingly the dispersion for the overlap areas and
for the intervening areas are nearly the same for the best shift!  
{\it The excess dispersion between the two light curves,
which is probably  due to  
microlensing, is removed}. 
At least from this
point of view the $\tau=415$ days solution is much more consistent than
the solution with $\tau=536$ days.

\begin{figure}[t]
\leavevmode\epsfbox{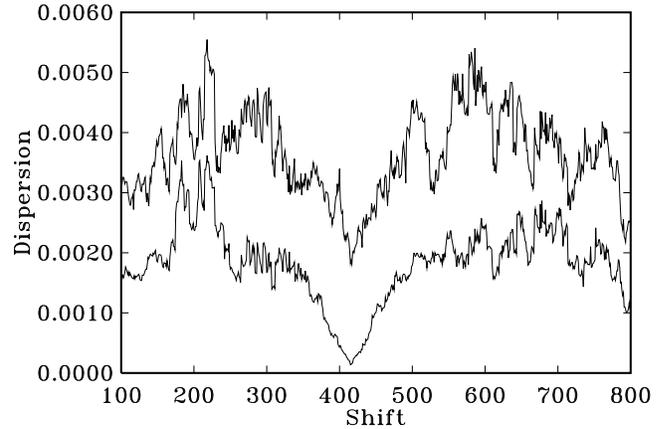}
\caption[ ]{Dispersion spectra for detrended
(with 9-degree polynomial) and artificial data sets.
The upper curve is $D^2_{\rm A,B}(\tau)$ for detrended data set and
the lower curve is $D^2_{\rm A,B}(\tau)$ for artificial data set.}
\label{fig9}
\end{figure}

We then repeated our experiment with
the 
artificial data set. In Fig.~\ref{fig9} the overall correspondence
between the two spectra is striking. It seems, however,
that the exact value of $415$ days 
for the very narrow dip of the real spectrum can be
slightly out of place, because it is ``cooked'' by high frequency local
fluctuation just
on the bottom of the feature. 

\begin{figure}[t]
\leavevmode\epsfbox{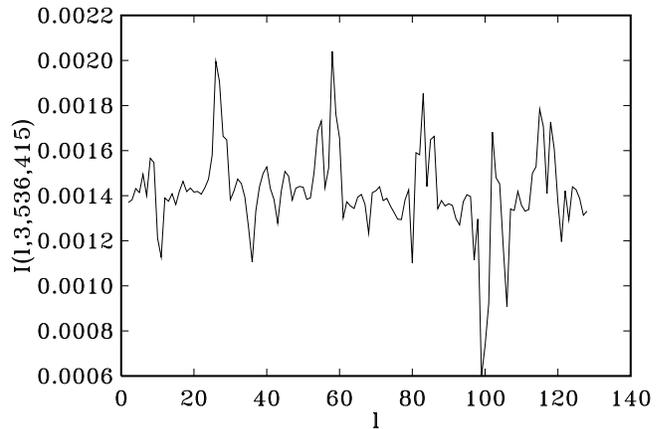}
\caption[ ]{Gain for hypothesis $\tau=536$ against hypothesis
$\tau=415$ for detrended data sets with three consecutive data
points removed.}
\label{fig10} 
\end{figure}

\begin{figure}[t]
\leavevmode\epsfbox{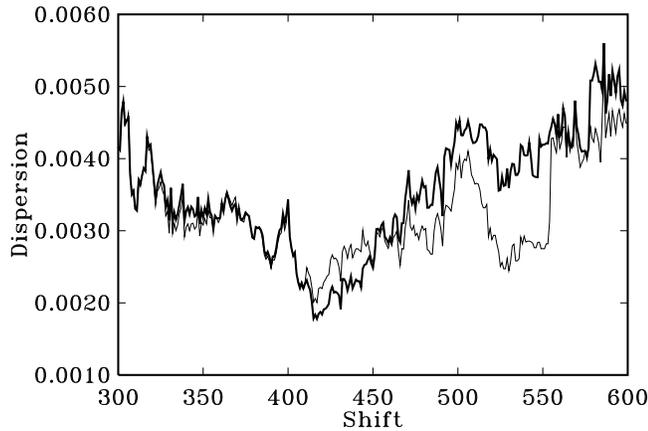}
\caption[ ]{Dispersion spectra for the two skipping schemes with
maximum contrast
between two hypotheses. The three point gaps started from
$l=58$ for the bold curve and from $l=99$ for the normal curve.}
\label{fig11}
\end{figure}

\begin{figure}[t]
\leavevmode\epsfbox{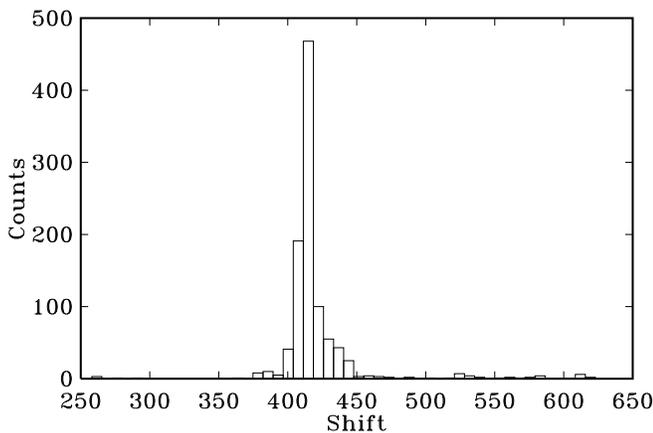}
\caption[ ]{Distribution of shifts for the bootstrap 
simulations for the optical
data.}
\label{fig12}
\end{figure}

The sensitivity analysis for the detrended data sets is also revealing.
In Figs.~\ref{fig10} and~\ref{fig11} it can be seen 
that the feature around $536$ days can be amplified
significantly or can be removed nearly completely by skipping three
consecutive observing nights from the original data sets. 
On the other hand the
depression around $415$ is rather stable. 
The bootstrap procedure gives us for the detrended data a standard error
of $\pm 32$ days for the
alternative solution of $415$ days. The distribution of
$1000$ trials is depicted in Fig.~\ref{fig12}. 

There are two competing values for the time delay, both of which we were
able to
derive from optical data using exactly the same techniques. We hoped
that the analysis of the radio data
could have given us clues as to which of them should be preferred.
\subsection{Radio data}
\begin{figure}[t]
\leavevmode\epsfbox{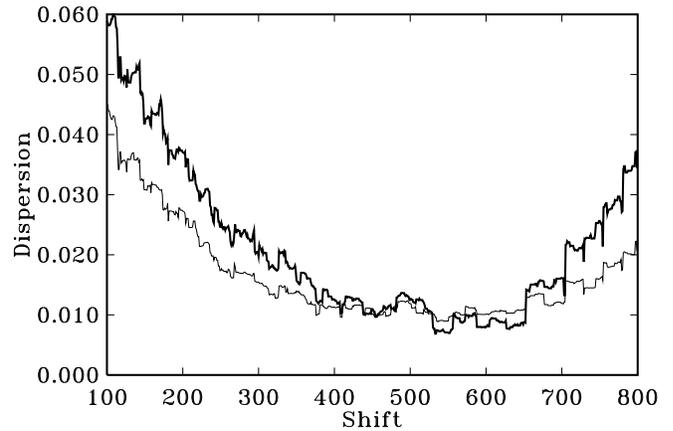}
\caption[ ]{Dispersion spectra for the original radio data sets.
The bold curve is $D^2_{\rm A,B}(\tau)$ and the normal curve is
$D^2_{\rm all}(\tau)$.}
\label{fig13}
\end{figure}

\begin{figure}[t]
\leavevmode\epsfbox{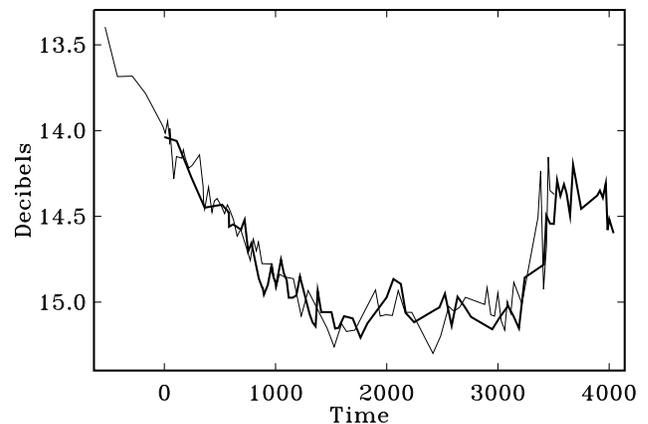}
\caption[ ]{Radio data sets 
combined for the best shift and parameter $a$.
The bold curve is A, the normal curve
is shifted B data.}
\label{fig14}
\end{figure}

We used the  radio data  published by Leh\'ar et al.~(\cite{Lehar92}) 
exactly in the manner they were used in PRHb. 
The observations marked in PRHb by an asterisk
were skipped and the other observed values were transformed
to logarithmic scale.
The $D_{\rm all}^2$ and $D_{\rm A,B}^2$ spectra (Eqs.~\ref{firststat} and
\ref{secondstat}) for 
the radio data are depicted in
Fig.~\ref{fig13}. The total minimum at ($533 \pm 40$) days (error
bars from bootstrapping)
is again in very good 
agreement with the result of PRHb. This coincidence
is so good that one may well be
inclined to drop the $\tau=415$
day 
hypothesis completely.
The plot of two light curves with appropriate shift and
parameter $a$ (Fig.~\ref{fig14}) fits well into this picture.

\begin{figure}[t]
\leavevmode\epsfbox{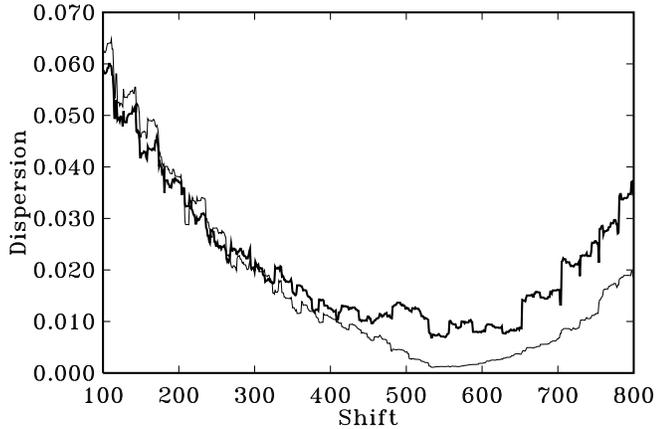}
\caption[ ]{Comparison of the dispersion spectra for the
original radio and artificial data sets. The bold curve is $D^2_{A,B}(\tau)$
for the original radio data, 
the normal curve is $D^2_{A,B}(\tau)$ for the artificial data.}
\label{fig15}
\end{figure}

Nonetheless,
we
applied to the radio data some of the tests already used for
the optical data.
We first repeated the experiments with artificial data sets generated
from the combined original data sets by 7-point median filtering. 
In Fig.~\ref{fig15} the plot of the original and the 
artificial spectra are shown. 
The results are
not very convincing, neither {\it pro} nor {\it contra}. However,
it seems that
the artificial spectrum (with $\tau=533$ days) has a slightly more pronounced 
minimum.

\begin{figure}[t]
\leavevmode\epsfbox{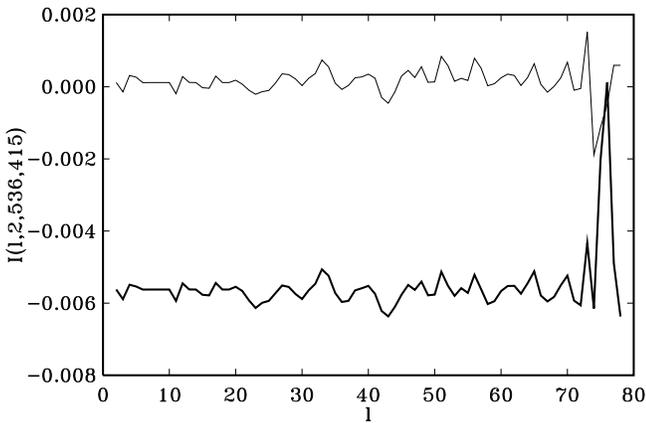}
\caption[ ]{Effect of data skipping to  $D_{A,B}^2(536)-D_{A,B}^2(415)$.
The bold curve depicts the gain (Eq.~\ref{gain}) computed for the 
original radio data sets. The
normal curve depicts the gain computed for 
the radio data with two skipped outlying points from B curve.}
\label{fig16}
\end{figure}

We then experimented with various skipping schemes, to look how
stable our results are against removal of a small number of observations.
It was quite surprising to find 
that the removal of {\it only two successive outlying
points from the B curve} can change the situation dramatically, see 
Fig.~\ref{fig16}.
The lower curve is the evaluation of 
the $\tau=536$ days hypothesis against the $\tau=415$
days hypothesis, when removing pairs of observations from {\it only 
the B curve}.
The upper curve depicts the same statistic for the data with 
two points removed from the B data set 
(the skipped points are from April 10, 1990 and May 7, 1990). 
It is seen that the data set with skips in the B curve
is relatively stable against further removing of time points. 
\begin{figure}[t]
\leavevmode\epsfbox{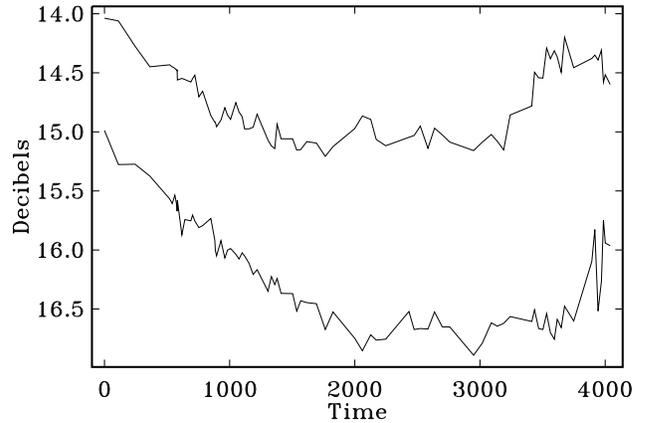}
\caption[ ]{Linearly interpolated observations of the A and B curve.
The upper curve is B, the lower curve is A.}
\label{fig17}
\end{figure}
In Fig.~\ref{fig17}
the original data points for the A and B curve are depicted, with linear
interpolation between the points. The ``bad'' points are exactly the 
points which cause the strong fluctuation at the very end of the B data
set.

\begin{figure}[t]
\leavevmode\epsfbox{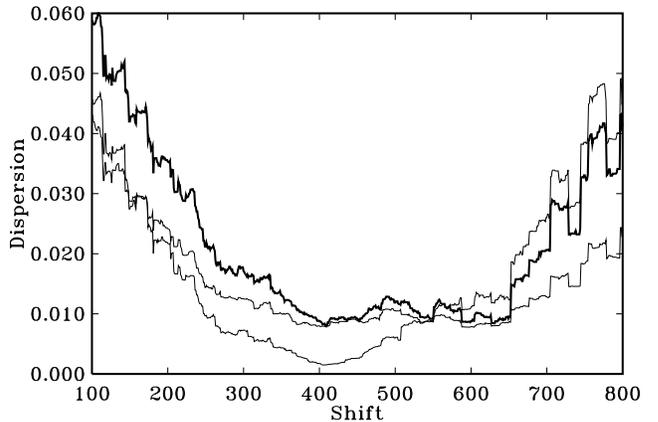}
\caption[ ]{Dispersion spectra for the radio data sets with skipped two
time
points from B curve.
The bold curve is $D^2_{\rm A,B}(\tau)$, the flat normal curve is 
$D^2_{\rm all}(\tau)$ for radio data with skips, the steeper normal curve
is the corresponding dispersion for artificial data.}
\label{fig18}
\end{figure}

\begin{figure}[t]
\leavevmode\epsfbox{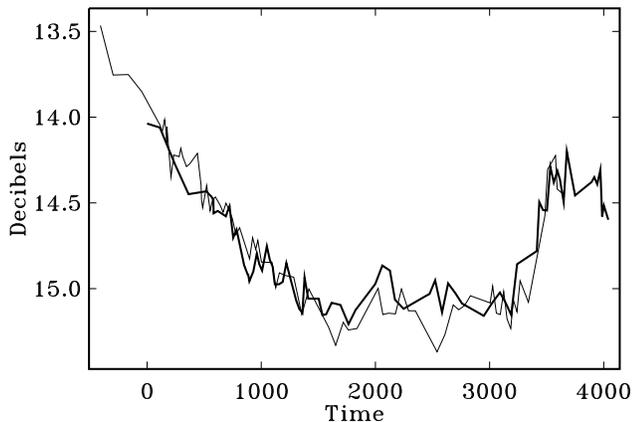}
\caption[ ]{Optimally shifted radio observations with two skips in B curve.
The bold
curve depicts A data and the normal curve shifted B data.}
\label{fig19}
\end{figure}

We proceed now with data set B without the two ``bad'' points. It is no
surprise to see that the main minimum of the $D_{\rm A,B}^2$ spectrum 
moves to $\tau=409$ days and that the overall correspondence between 
the artificial
spectrum and the real one is also good (see Fig.~\ref{fig18}). 
Quite pleasing
is also the plot of the optimally shifted B curve with two skipped
points and the original A
curve (see Fig.~\ref{fig19}). 

Bootstrapping gives us an error
of $\pm 23$ days
for the shift, thus the correspondence between 
the results obtained from the
optical
and the radio observations is rather good. 
The error value must not be taken too seriously, since the
use of a 7-point median filtering to build the mean curve
estimates is somewhat arbitrary. However, it
can be understood as a reasonable guess.
\section{Discussion}
Let us first take the original optical and radio data sets as they are. We
were able to reproduce the results of the rather complex
analysis in PRHa and PRHb, using  
extremely simple statistics.
However, the simple formulation of our procedures also allowed to get
additional insights into what is actually going on. We now
summarize the {\it pro} and {\it contra} arguments.
\subsection{{\em Contra}  $\tau=536$ days}
The feature around $\tau=536$ days is located near the absolute minimum
of the window function in the optical dispersion
spectra and does not have a profile
which is expected from the artificial curves. 
{\em There is a large excess for the dispersion 
in the overlapped regions of the shifted light curves, and this is
revealed using both methods of analysis.} The last fact also invalidates
the Monte-Carlo results in PRHa because their data set
models are too regular compared to the real data sets. 
This is well illustrated in 
Fig. ~\ref{fig2}.
The dispersion spectra in the $536$ days region are rather sensitive to
skipping few data points and detrending.
\subsection{{\em Pro}  $\tau=536$ days}
Original data sets are used. The statistically sound optimal prediction
method and simple explorative type analysis method give essentially the
same results. No additional or excessively metaphysical arguments are
involved in the data analysis. Good 
correspondence between optical and radio results.
\subsection{{\em Contra}  $\tau=415$ days}
The good looking results are obtained after somewhat arbitrary detrending
and removing part of the data from analysis. 
From the purist's point of view this is unacceptable.
\subsection{{\em Pro}  $\tau=415$ days}
For both data sets the results of the explorative analysis 
are relatively consistent and stable.
The dispersion for the overlapped region in the spectrum for the optical
data is nearly the same as for other regions. The overall form
of the depressions in the dispersion spectra mimic quite well model spectra.
\subsection{Future work}
After the first version of this paper was completed a more extensive
optical data set became available to us. These data have been
obtained by R.~Schild and contain 707 time points
from November 1979 to June 1993.   
Here we present only a very short analysis of the new
data, details will be presented elsewhere. 
\begin{figure}[t]
\leavevmode\epsfbox{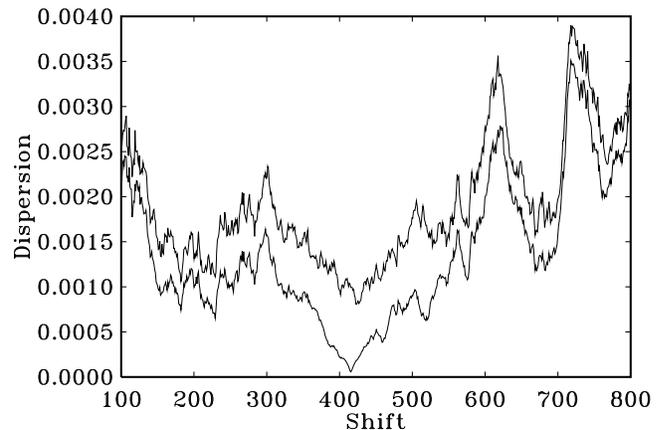}
\caption[ ]{Dispersion spectra for the extended data set 
provided to us by R.~Schild.
The upper curve is $D^2_{\rm A,B}(\tau)$, the lower curve is
$D^2_{\rm A,B}(\tau)$ for the corresponding artificial data set with
time shift $\tau=415$ days.}
\label{fig21}
\end{figure}
In Fig.~\ref{fig21}  the dispersion spectrum for the full
data set {\it without any detrending and without any skips} is plotted against
the spectrum of the
artificial data set which was computed by 7-point median
filtering of the coalesced original data (with B curve shifted by $415$ days).
It can be seen that the feature around $536$ days is practically absent
and the depression around $415$ days is well pronounced. This result is
in good agreement with the analysis made by Schild and Thomson 
(\cite{Schild93}) and Thomson and Schild (\cite{Thomson}).

The number of observations in the extended 
data set is now much larger and probably sufficient to address the rather
complicated question about the statistical significance of the
feature around 415 days. As we demonstrated in Section 2
the straightforward Monte-Carlo modelling can lead us to a wrong conclusion.
Therefore we plan to use more complex waveform models for the
data sets which will include
also parts describing microlensing. 
\section{Conclusions}
We introduced an extremely simple, but nevertheless sensitive method
to seek probable time delay values for light curves which are
assumed to originate from one and the same source. The method
easily recovered results which were obtained by more complex and
statistically sound procedures presented in PRHa and PRHb. However,
the simplicity of the proposed method allowed us to get additional
insights into the problems which originate from unfortunate
spacing of the sampling points for the optical light curves. It allowed
us also to produce an 
alternative consistent solution for the time delay problem.

The dispersion spectra of the optical and radio data can show minima
in various places depending on the method of analysis and
preprocessing of the data.
For theoretical reasons delays $\tau \leq 0$ can be excluded. 
The minimum
near 1.5 years most probably
results from a statistical fluctuation which is amplified 
due to the windowing, since for this time shift the overlap
of data from the two light curves is at a minimum and thus the least
squares minimization has more freedom to fit free
parameters. 

For the radio data we found that removing only two observations
from B 
gives a stable spectrum for the dispersions, with a best value for the 
shift near $\tau=409$ days. This result is in good agreement with
the value of $\tau=415$ days for the 
optical curve, which reveals itself after the 
detrending of the original data, or in other circumstances which
allow to depress the instable minimum around $536$ days.

We think that the time
delay controversy on QSO 0957+561 A,B is still not settled, but there
is quite strong evidence (especially if we take into account 
preliminary results
obtained using the extended data set by R. Schild) that the 
time delay
is in the region of 400-420 days.

\begin{acknowledgements}
We wish to thank L.~Nieser, P.~King and A.~Dent for helpful
discussions and computational aid. It is a pleasure to
acknowledge valuable discussion with Rudy Schild and David Thomson,
who also kindly provided us with their unpublished data. 
We finally appreciate the 
comments and constructive 
criticism on the first version of this paper by an anonymous referee.
Part of this work was supported by the
      {\it Deut\-sche For\-schungs\-ge\-mein\-schaft, DFG\/} under
      Re 439, Schr 417 and 436 EST.
\end{acknowledgements}
\appendix
\section{Appendix. Nonparametric dispersion estimates}
\subsection{Dispersion estimation through subsampling}
For a set of variables
$y_i, i=1,2,\dots,N$ we observe that 
\begin{eqnarray}
& &{1\over 2N^2}\sum_{i=1}^N\sum_{j=1}^N (y_i-y_j)^2
= \nonumber \\
& = &{1\over 2N^2}\sum_{i=1}^N\sum_{j=1}^N 
\big((y_i-\mu)-(y_j-\mu)\big)^2= \nonumber \\
& = &{1\over N}\sum_{i=1}^N(y_i-\mu)^2-{1\over N^2}\big(\sum_{i=1}^N(y_i-\mu)
\big)^2.
\end{eqnarray}
If we set $\mu={1\over N}\sum_{i=1}^N y_i$ the second term in the 
last expression
vanishes and consequently the first double sum can be considered as 
a standard 
(slightly biased) estimator for the sample dispersion. Taking into
account that $(y_i-y_j)^2=(y_j-y_i)^2$ and using another normalization
we can obtain the standard unbiased estimate $S^2$ for the dispersion:
\begin{eqnarray}
\label{cross}
S^2 & = &{1\over N(N-1)}\sum_{i=1}^{N-1}\sum_{j=i+1}^N(y_i-y_j)^2=
\nonumber \\
 & = &{1\over N-1}\sum_{i=1}^N(y_i-\mu)^2.
\end{eqnarray}
Let $K={N(N-1)/ 2}$ be the total number of squares in the first sum, so
that:
\begin{equation}
S^2={1\over K}\sum_{k=1}^K{(y_{i(k)}-y_{j(k)})^2\over 2}
={1\over K}\sum_{k=1}^K d_k.
\end{equation}
The sequence of half squares $d_k={(y_{i(k)}-y_{j(k)})^2 / 2}$ 
can be looked upon as a
general finite sample from which we can draw random subsamples 
$d_{k(l)},l=1,2,...,L$. In this
case the mean values
\begin{equation}
{\hat S}^2={1\over L}\sum_{l=1}^L d_{k(l)}
\end{equation}
approximate the original estimate $S^2$. How good are these approximations?
From standard sampling theory (Cochran \cite{Cochran})
we get the 
estimate for the dispersion
of ${\hat S}^2$ ({\it subsampling dispersion}):
\begin{equation}
{\bf D}({\hat S}^2)={\bf E}(\hat S^2-{\bf E}\hat S^2)={(K-L)D^2\over KL},   
\end{equation}
where {\bf E} is mathematical expectation operator, 
$D^2$ is the dispersion for the full sample of half squares
\begin{equation}
D^2={\sum_{k=1}^K (d_k-\bar d)^2 \over K-1},
\end{equation}
and $\bar d$ is their mean value. We see that 
the approximation error vanishes if $L$ approaches $K$. 
The important point is that
even for small values of $L$ the 
subsampling error is as small as $D^2/L$.

For the particular
case of normally distributed $y_i$s the half squares are distributed
according to the scaled ${\chi}^2(1)$ law and consequently $D^2$ can
be substituted by the value of $2\sigma^2$ where $\sigma$ is the standard
deviation for the $y_i$ or some of its estimates.  

Putting now all things
together we get the main result of this paragraph: the dispersion 
$S^2$ can be estimated as {\it half sum of squares chosen randomly
from the full set}. The approximate 
dispersion of this estimate can generally be
computed from any available estimate of the original dispersion of 
$y_i$, and in
particular from ${\hat S}^2$ itself:
\begin{equation}
{\bf D}({\hat S}^2)={2(K-L){\hat S}^2\over KL}.
\end{equation} 
Of course, for a simple estimation of the dispersion, our scheme is rather
cumbersome, but as we will see later, it can be very useful in more
complicated situations. In mathematical statistics the estimates of type
(\ref{cross}) are known as {\it U--statistics} (Hoeffding, \cite{Hoeffding}) 
and estimates
based on subsampling as {\it incomplete U--statistics} (Blom, \cite{Blom}).
\subsection{Trend as nuisance}
Let us now consider the regression model
\begin{equation}
y_i=g(t_i)+\epsilon_i=g_i+\epsilon_i, i=1,2,...,N,
\end{equation}
where the $t_i$ are randomly distributed time points, $g(t)$ is some smooth
but unknown function of time (trend), the $y_i$ are observed values and
the 
$\epsilon_i$ are unknown observational errors.
We are seeking an estimate for the dispersion of the $\epsilon_i$.

Normally the trend  $g(t)$ is approximated by some parametric
model and estimates for the $\epsilon_i$ and also for their 
dispersion are computed 
from residuals after
some fitting procedure. But, as demonstrated below, the estimate for
the dispersion can be computed without knowing the exact form of $g(t)$.

Let us first write down the general dispersion of the observed values
\begin{eqnarray}
\label{trend}
S_{\rm obs}^2 & = &{1\over
N(N-1)}\sum_{i=1}^{N-1}\sum_{j=i+1}^N(y_i-y_j)^2= \nonumber \\
& = &{1\over N(N-1)}\sum_{i=1}^{N-1}
\sum_{j=i+1}^N((e_i-e_j)+(g_i-g_j))^2= \nonumber \\
& = &S_{\rm err}^2+S_{\rm trend}^2 \nonumber \\
& & +{2\over N(N-1)}\sum_{i=1}^{N-1}\sum_{j=i+1}^N(e_i-e_j)(g_i-g_j).
\end{eqnarray}
The last sum can be considered as small (at least for samples containing
enough observations) 
since trend and error are not correlated.
The value for the trend dispersion 
$S_{\rm trend}^2$ can, however, be significant.
In order to get an estimate for the error dispersion 
$S_{\rm err}^2$ we must somehow eliminate
the trend dispersion from the general sum. From the discussion above we know
that we are not forced to use all pairs of squares to get dispersion estimates.
Let us choose only those pairs which have close enough observing times.
Formally we can define a {\it selection window}
\begin{equation}
G(\delta,t_i,t_j)=\cases{$1$ &$\vert t_i-t_j \vert \leq 
\delta$\cr
$0$ & otherwise\cr}
\end{equation}
for some specified maximum allowable lag $\delta$, and enter it into
the general sum
\begin{equation}
{\hat S}^2(\delta)={1\over 2}{\sum_{i=1}^{N-1}
\sum_{j=i+1}^NG(\delta,t_i,t_j)(y_i-y_j)^2
\over
\sum_{i=1}^{N-1}\sum_{j=i+1}^N G(\delta,t_i,t_j)}.
\end{equation}
This choice of a 
subsample from the total of squares is not random from the point
of view of its construction, but it is random from the point of view of
the statistics of errors. 

For a slowly varying trend $g(t)$ we can assume
that it is Lipschitz continuous with some maximum slope parameter~$A$
\begin{equation}
\vert g(t_i)-g(t_j) \vert \leq A \vert t_i -t_j \vert
\end{equation}
and consequently the upper limit for the dispersion of the trend part
can be estimated by
\begin{eqnarray}
& &{1\over 2}{\sum_{i=1}^{N-1}\sum_{j=i+1}^NG(\delta,t_i,t_j)(g_i-g_j)^2
\over
\sum_{i=1}^{N-1}\sum_{j=i+1}^N G(\delta,t_i,t_j)}\leq \nonumber \\
& \leq &{1\over 2}{\sum_{i=1}^{N-1}\sum_{j=i+1}^N
G(\delta,t_i,t_j)A^2(t_i-t_j)^2
\over
\sum_{i=1}^{N-1}\sum_{j=i+1}^N G(\delta,t_i,t_j)}\leq \nonumber \\
& \leq & {1\over 2}(A\delta)^2
\end{eqnarray}
The last limit is quite conservative, so that it should be used only for
rather crude guesses. Fortunately, the second sum can also be computed
from the time point sequence. 

For every fixed value of $\delta$ there are 
\begin{equation}
L=\sum_{i=1}^{N-1}\sum_{j=i+1}^N G(\delta,t_i,t_j)
\end{equation} 
pairs of squares to
be taken into account when estimating the dispersion. 
The error from subsampling
increases with decreasing $L$ (and of course with decreasing 
$\delta$). But the
bias due to the trend part of the observed values 
({\it modelling error}) decreases with decreasing
$\delta$. Consequently there is a certain value for $\delta$ where both errors
are of comparable magnitude. Let us call this tradeoff value {\em optimal}
(see Fig. \ref{fig4}). 
The 
corresponding ${\hat S}^2(\delta)$ can be used as an estimate for
the original error dispersion.

There is another well known procedure to choose a subsample of 
squared differences.
We can select only pairs which consists of neighbouring observations
in the original data. The corresponding estimate for dispersion is now
\begin{equation}
{\hat S}^2={1\over 2(N-1)}\sum_{i=1}^{N-1}
(y_i-y_{i+1})^2.
\end{equation}
In this estimate we avoided introduction of any
free parameters into the selection scheme.  
Similar estimates were proposed
by different authors (see von Neumann et al.~\cite{Neumann} or in the
context of periodicity search Lafler\&Kinman \cite{Lafler}).

It is also possible to combine two subsampling schemes given
above. In this case only 
neighbouring
pairs of observations $y_i,y_j$  for which
$\vert t_i-t_j \vert \leq \delta$ are included into the subsample.
Thereby we avoid additional scatter in our estimates 
due to the {\it long gaps} in the data set. However, every additional
restriction for selected pairs decreases their total number
and consequently increases the subsampling error.

\end{document}